\documentclass{ametsocV5}

\usepackage{amsmath,amsfonts,amssymb,bm}
\usepackage{mathptmx}
\usepackage{newtxtext}
\usepackage{newtxmath}
\usepackage{enumitem}
\usepackage{soul}

\title{LiSBOA: LiDAR Statistical Barnes Objective Analysis for optimal design of LiDAR scans and retrieval of wind statistics.  Part I: Theoretical framework}

  \authors{Stefano Letizia, Lu Zhan and Giacomo Valerio Iungo\correspondingauthor{Giacomo Valerio Iungo, 
      valerio.iungo@utdallas.edu}}

     \affiliation{Wind Fluids and Experiments (WindFluX) Laboratory, Mechanical Engineering Department, The University of Texas at Dallas}

\abstract{A LiDAR Statistical Barnes Objective Analysis (LiSBOA) for optimal design of LiDAR scans and retrieval of the velocity statistical moments is proposed. The LiSBOA represents an adaptation of the classical Barnes scheme for the statistical analysis of unstructured experimental data in $N$-dimensional spaces and it is a suitable technique for the evaluation over a structured Cartesian grid of the statistics of scalar fields sampled through scanning LiDARs. The LiSBOA is validated and characterized via a Monte Carlo approach applied to a synthetic velocity field. This revisited theoretical framework for the Barnes objective analysis enables the formulation of guidelines for optimal design of LiDAR experiments and efficient application of the LiSBOA for the post-processing of LiDAR measurements. The optimal design of LiDAR scans is formulated as a two cost-function optimization problem including the minimization of the percentage of the measurement volume not sampled with adequate spatial resolution and the minimization of the error on the mean of the velocity field. The optimal design of the LiDAR scans also guides the selection of the smoothing parameter and the total number of iterations to use for the Barnes scheme.} 

\begin{document}

\maketitle

%
%
%

%
\section{Introduction}
Reliable measurements of the wind-velocity vector field are essential to understand the complex nature of atmospheric turbulence and provide valuable datasets for the validation of theoretical and numerical models. However, field measurements of wind speed are typically characterized by large uncertainties due to the generally unknown and uncontrollable boundary conditions \citep{Braham1979}, the broad range of time and length-scales \citep{Roisin2010_scales}, and the complexity of the physics involved \citep{Stull1988_turb}. Furthermore, the large measurement volume, which typically extends throughout the height of the atmospheric boundary layer, imposes to the experimentalists the selection of the sampling parameters as a trade-off between spatial and temporal resolutions.

Wind speed has been traditionally measured through local sensors, such as mechanical, sonic, and hot-wire anemometers \citep{Liu2019,Kunkel2006}. Besides their simplicity, mechanical anemometers are affected by errors due to the flow distortion of the supporting structures and harsh weather conditions (e.g. ice) \citep{Mortensen1994}. Furthermore, their relatively slow response results in a limited range of the measurable time-length scales, which makes them unsuitable, for instance, to measure the turbulent flow around urban areas \citep{Pardyjak2017}. Sonic anemometers can measure the three velocity component with frequencies up to 100 Hz \citep{Cuerva2000} in a probing volume of the order of 0.01 m, yet measurements might be still affected by the wakes generated by the supporting structures, such as met-towers and struts and they are sensitive to temperature variations \citep{Mortensen1994}. Hot-wire anemometers, although they provide a full characterization of the energy spectrum, require a complicated calibration \citep{Kunkel2006} and are extremely fragile \citep{Wheeler2004}.
Furthermore, traditional single-point sensors are unable to provide an adequate characterization of the spatial gradients of the wind velocity vector, which is particularly significant in the vertical direction \citep{Roisin2010_strat}. To overcome this issue, several anemometers arranged in arrays and supported by meteorological masts have been deployed in several field campaigns \citep{Haugen1971,Bradley1983,Taylor1987,Emeis1995,Pashow2001,Berg2011,Kunkel2006}.

In the last few decades, remote sensing instruments have been increasingly utilized to probe the atmospheric boundary layer \citep{Debnath2017,Debnath2017b} and nowadays they represent a more cost-effective and flexible alternative to meteorological towers \citep{Newsom2017}. In particular, in the realm of remote sensing anemometry, Doppler wind light detection and ranging (LiDAR) systems underwent a rapid development due to the significant advancement in eye-safe laser technology \cite{Emeis2010}. Wind LiDARs have been heavily employed in wind energy \citep{Bingol2010,Aitken2014,Trujillo2011,IungoJTECH2013,Machefaux2016,Garcia2017,ElAsha2017,Bromm2018,Zhan2019,Zhan2020}, airport monitoring \citep{Kopp2005,Tang2011,Holzapfel2016,Thobois2019}, micro-meteorology \citep{GalChen1992,Banakh1998,Banta2006,Mann2010,Munoz2012,Rajewski2013,Schween2014}, urban wind research \citep{Davies2007,Newsom2008,Xia2008,Kongara2012,Huang2017,Halios2018} and studies of terrain-induced effects\citep{Bingol2009,Krishnamurthy2013,Kim2016,Pauscher2016,Risan2018,Fernando2019,Bell2020}.

Besides the mentioned capabilities, LiDARs present some important limitations, such as reduced range in adverse weather conditions (precipitations, heavy rain or low aerosol concentration) \citep{Liu2019} and a limited spatio-temporal resolution of this instrument, namely about 20 meters in the radial direction and about 10 Hz in sampling frequency. These technical specifications, associated with the non-stationary wind conditions typically encountered for field experiments, pose major challenges to apply wind LiDARs for the statistical analysis of turbulent atmospheric flows. 

In the realm of wind energy, early LiDAR measurements were limited to the qualitative analysis of snapshots of the line-of-sight (LOS) velocity, i.e. the velocity component parallel to the laser beam \citep{Kasler2010,Clive2011}. Fitting of the wake velocity deficit was also successfully exploited to extract quantitative information about wake evolution from LiDAR measurements \citep{Aitken2014,Wang2015,Kumer2015,Trujillo2016,Bodini2017}. To characterize velocity fields with higher statistical significance, the time averages of several LiDAR scans were calculated for time periods with reasonably steady inflow conditions \citep{IungoJTECH2014, Machefaux2016, VanDooren2016}. In the case of data collected under different wind and atmospheric conditions, clustering and bin-averaging of LiDAR data were carried out \citep{Machefaux2016,Garcia2017,Bromm2018, Zhan2019, Zhan2020}. Finally, more advanced techniques for first-order statistical analysis, such as variational methods \citep{Xia2008,Newsom2004}, optimal interpolation \citep{Xu2002,Kongara2012}, least-squares methods \citep{Newsom2008}, Navier-Stokes solvers \citep{LINCOM,Sekar2018} were applied for the reconstruction of the velocity vector field from dual-Doppler measurements.

Besides the mean field, the calculation of higher-order statistics from LiDAR data to investigate atmospheric turbulence is still an open problem. In this regard, \cite{Eberhard1989} re-adapted the post-processing of the velocity azimuth display (VAD) scans \citep{Lhermitte1969, Wilson1970, Kropfli1985} to estimate all the components of the Reynolds stress tensor by assuming horizontal homogeneity of the flow within the scanning volume, which can be a limiting constraint for measurements in complex terrains \citep{Bingol2009, Frisch1991}. Range height indicator (RHI) scans were used to detect second-order statistics \citep{Bonin2017}, spectra, skewness, dissipation rate of the velocity field, and even heat flux \citep{GalChen1992}. 

A typical scanning strategy to obtain high-frequency LiDAR data consists in performing scans with fixed elevation and azimuthal angles of the laser beam while maximizing the sampling frequency \citep{Mayor1997,OConnor2010, Vakkari2017, Frehlich2002,Debnath2017,Choukulkar2017,Lundquist2017}. Recently, in the context of wind radar technology, but readily applicable to LiDARs as well, a promising method for the estimation of the instantaneous turbulence intensity based on the Taylor hypothesis of frozen turbulence was proposed by \cite{Duncan2019}. More advanced techniques exploit additional information of turbulence carried by the spectrum of the back-scattered LiDAR signal \citep{Smalikho1995}. However, this approach requires the availability of LiDAR raw data, which is not generally granted for commercial LiDARs. For a review on turbulence statistical analyses through LiDAR measurements, the reader can refer to \cite{Sathe2013}. 

For remote sensing instruments, data are typically collected based on a spherical coordinate system, then interpolated over a Cartesian reference frame oriented with the $x$-axis in the mean wind-direction. This interpolation can be a source of error \citep{Fuertes2018b}, especially if a linear interpolation method is used \citep{Garcia2017,Fuertes2018,Beck2017,LINCOM}. Delaunay triangulation has also been widely adopted for coordinate transformation \citep{Clive2011,Trujillo2011,IungoJTECH2014,Trujillo2016,Machefaux2016}, yet with accuracy not quantified in case of non-uniformly distributed data. It is reasonable to weight the influence of the experimental points on their statistics by the distance from the respective grid centroid, such as using uniform \citep{Newsom2008}, hyperbolic \citep{VanDooren2016} or Gaussian weights \citep{Newsom2014,Wang2015,Zhan2019}. The use of distance-based Gaussian weights for the interpolation of scattered data over a Cartesian grid is at the base of the Barnes objective analysis (or Barnes scheme) \citep{Barnes1964}, which has been extensively used in meteorology. It represents an iterative statistical ensemble procedure to reconstruct a scalar field arbitrarily sampled in space and low-pass filtered with a cut-off wavelength that is a function of the parameters of the scheme.

The scope of this work is to define a methodology to post-process scattered data of a turbulent velocity field measured through a scanning Doppler wind LiDAR to calculate mean, standard deviation and even higher-order statistical moments  on a Cartesian  grid. The proposed methodology, referred to as LiDAR Statistical Barnes Objective Analysis (LiSBOA), represents an adaptation of the classic Barnes scheme to $N$-dimensional domains enabling applications for non-isotropic scalar fields through a coordinate transformation. A major point of novelty of the LiSBOA is the estimation of wind-velocity variance (and eventually higher-order statistics) from the residual field of the mean, which also provides adequate filtering of dispersive stresses due to data variability not connected with the turbulent motion. A criterion for rejection of statistical data affected by aliasing due to the undersampling of the spatial wavelengths under investigation is formulated. The LiSBOA is assessed against a synthetic scalar field to validate its theoretical response and the formulated error metric. Finally, detailed guidelines for the optimal design of a LiDAR experiment and effective reconstruction of the wind statistics are provided. 

The remainder of the manuscript is organized as follows: in section \ref{sec:Theory} the extension of the Barnes scheme theory to $N$-dimensional domains and higher-order statistical moments is presented. In section \ref{sec:MonteCarlo}, the theoretical response function of the LiSBOA is validated against a synthetic case, while guidelines for proper use of the proposed algorithm are provided in section \ref{sec:Guidelines}. Finally, concluding remarks are provided in section \ref{sec:Conclusions}. 

\section{The Barnes Objective Analysis: fundamentals and extension to statistical $N$-dimensional analysis}\label{sec:Theory}
The Barnes scheme was originally conceived as an iterative algorithm aiming to interpolate a set of sparse data over a Cartesian grid \citep{Barnes1964} and it was inspired by the successive correction scheme by \cite{Cressman1959}. The first iteration of the algorithm calculates a weighted-space-averaged field, $g^0$, over a Cartesian grid from the sampled scalar field, $f$. The mean field is iteratively modified by adding contributions to recover features characterized by shorter wavelengths, which are inevitably damped by the initial averaging process. In this work, we adopt the most classical form of the Barnes scheme as follows:
\begin{equation}\label{eqn:Barne_scheme}
\begin{cases}
	g^0_i = \sum\limits_{j} w_{ij} f_j\\
	g^m_i=\sum\limits_{j} w_{ij} (f_j - \phi(g^{m-1})_j)+g_i^{m-1} \quad \forall \quad m\in \mathbb{N}^+,
\end{cases}
\end{equation}
where $g^m_i$ is the average field at the $i$-th grid node with coordinates $\mathbf{r_i}$ for the $m$-th iteration, $f_j$ is the scalar field sampled at the location $\mathbf{r_j}$ and $\phi$ represents the linear interpolation operator from the Cartesian grid to the sample location. The weights for the sample acquired at the location $\mathbf{r_j}$ and for the calculation of the statistics of $f$ at the grid node with coordinates $\mathbf{r_i}$, $w_{ij}$, are defined as:
\begin{equation}\label{eqn:Barne_weights}
	w_{ij}= \frac{e^{-\frac{|\mathbf{r_i}-\mathbf{r_j}|^2}{2 \sigma^2}}}{\sum\limits_{j} e^{-\frac{|\mathbf{r_i}-\mathbf{r_j}|^2}{2 \sigma^2} }},
\end{equation}
where $\sigma$ is referred to as smoothing parameter and $|.|$ indicates Euclidean norm. For practical reasons, the summations over $j$ are performed over the neighboring points included in a ball with a finite radius $R_{\text{max}}$ (also called the radius of influence) and centered at the $i$-th grid point. In this work, following \cite{Barnes1964}, we selected $R_{\text{max}}=3 \sigma$, which encompasses 99.7\% and 97\% of the volume of the weighting function in 2D and 3D, respectively.

In literature, there is a lack of consensus for the selection of the total number of iterations \citep{Barnes1964,Achtemeier1989,Smith1984,Seaman1989} and the smoothing parameter \citep{Barnes1994a,Caracena1987,Pauley1990}. A reduction of the smoothing parameter, $\sigma$, as a function of the iteration, $m$, was originally proposed by \cite{Barnes1973}; however, this approach resulted to be detrimental in terms of noise suppression \citep{Barnes1994c}.

In the frequency domain, the Barnes objective analysis is tractable as a low-pass filter applied to a scalar field, $f$, with a response as a function of the spatial wavelength depending on the smoothing parameter, $\sigma$, and the number of iterations, $m$. This feature has been exploited in meteorology to separate small-scale from mesoscale motions \citep{Doswell1977, Maddox1980, Gomis1990}. The spectral behavior of the Barnes scheme has been traditionally characterized by calculating the so-called continuous response at the $m$-th iteration, $D^m(\mathbf{k})$, with $\mathbf{k}$ being the wavenumber vector. $D^m(\mathbf{k})$ is defined as the ratio between the amplitude of the Fourier mode $e^{\mathrm{i} \mathbf{k}\cdot \mathbf{x}}$ (with $\mathrm{i}=\sqrt{-1}$) for the reconstructed field, $g^m$, to its amplitude for in input field, $f$, in the limit of a continuous distribution of samples and infinite domain. The analytical expression for the continuous response was provided by \cite{Barnes1964} and \cite{Pauley1990} for 1D and 2D domains, respectively, while in the context of the LiSBOA it is extended to $N$ dimensions to enhance its applicability. Furthermore, besides the spatial variability of $f$, the temporal coordinate, $t$, is introduced to determine the response of the statistical moments of $f$. 
Although the present discussion hinges on the choice of time as a non-spatial variable, the same approach can be applied to other non-spatial variables, such as Obukhov length to characterize atmospheric stability, operative conditions of a wind turbine, wind direction, assuming that statistical homogeneity holds along that coordinate.

We consider a continuous scalar field, $f(\textbf{x},t)$, which is defined over an $N$-dimensional domain, $\textbf{x}$. It is further assumed that the field $f$ is ergodic in time. By adopting the approach proposed by \cite{Pauley1990} and by taking advantage of the isotropy of the Gaussian weights (Eq. (\ref{eqn:Barne_weights})), we can define the LiSBOA operator at the 0-th iteration as:
\begin{equation}\label{eqn:Theory_Barnes}
g^0(\textbf{x})=\frac{1}{(\sqrt{2\pi}\sigma)^N} \int_{\mathbb{R}^N} \left[ \frac{1}{t_{2} -t_{1}} \int_{t_{1}} ^{t_{2}} f(\bm{\xi},t) dt \right] e^{-\frac{|\mathbf{x}-\bm{\xi}|^2}{2 \sigma^2}} d \bm{\xi},
\end{equation}
where $t_1$ and $t_2$ are initial and final time. The term within the square brackets represents the mean of $f$ over the considered sampling interval $[t_1,~t_2]$, which is indicated as $\overline{f}$. Moreover, to reconstruct a generic $q$-th central statistical moment of the scalar field, $f$, it is sufficient to apply the LiSBOA operator of Eq. (\ref{eqn:Theory_Barnes}) to the fluctuations over $\overline{f}$ to the $q$-th power:
\begin{equation}\label{eqn:Theory_Barnes_moments}
\mu^{q}_f(\mathbf{x}) =\frac{1}{(\sqrt{2\pi}\sigma)^N} \int_{\mathbb{R}^N} \left\{ \frac{1}{t_{2} -t_{1}} \int_{t_{1}} ^{t_{2}} \left[f(\bm{\xi},t)-\overline{f}(\bm{\xi},t)\right]^q dt \right\} e^{-\frac{|\mathbf{x}-\bm{\xi}|^2}{2 \sigma^2}} d \bm{\xi}.
\end{equation}
For practical applications, the mean field $\overline{f}$ is generally not known, but it can be approximated by the LiSBOA output, $g^m$, interpolated at the sample location through the operator $\phi$. By comparing Eq. (\ref{eqn:Theory_Barnes_moments}) with Eq. (\ref{eqn:Theory_Barnes}), it is understandable that the response function of any central moment with order higher than one is equal to that of the $0$-th iteration response of the mean, $g^0$. Indeed, Eq. (\ref{eqn:Theory_Barnes}) can be interpreted as the $0$-th iteration of the LiSBOA spatial operator (viz. Eq. (\ref{eqn:Theory_Barnes})) applied to the fluctuation field to the $q$-th power.

By leveraging the convolution theorem, it is possible to calculate the response function of the mean of the $0$-th iteration of the LiSBOA in the frequency domain (see  Appendix A for more details). This result, combined with the recursive formula of \cite{Barnes1964} for the response at the generic iteration $m$, provides the spectral response of the LiSBOA for the mean:
\begin{equation}
D^m= 
\begin{cases}
 D^0(\textbf{k})=e^{-\frac{\sigma^2}{2} |\mathbf{k}|^2}=e^{-\frac{\sigma^2\pi^2}{2} \left[\sum_{p=1}^N\frac{1}{\Delta n_p^2}\right]} ~~~~~~~~~~~~~~~~\text{for}  ~ m=0 \\
 D^0 \sum_{p=0}^m (1-D^0)^p ~~~~~~~~~~~~~~~~~~~~~~~~~~~~~~~~~~~~~~~~~~~~\text{for}  ~ m\in \mathbb{N}^+, 
\end{cases}
\label{eqn:Barnes_response}
\end{equation}
where ${\mathbf{\upDelta} \mathbf{n}}$  is the half-wavelength vector associated with $\mathbf{k}$. Equation \ref{eqn:Barnes_response} states that, for a given wavenumber (i.e. half-wavelength), the respective amplitude of the interpolated scalar field, $g^m$, is equal to that of the original scalar field, damped with a function of the smoothing parameter, $\sigma$, and the number of iterations, $m$. This implies that the parameters $\sigma$ and $m$ should be selected properly to avoid significant damping for wavelengths of interest or dominating the spatial variability of the scalar field under investigation.

For real applications, the actual LiSBOA response function can depart from the above-mentioned theoretical response (Eq. (\ref{eqn:Barnes_response})) for the following reasons:
\begin{itemize}
\item the convolution integral in Eq. (\ref{eqn:Theory_Barnes}) is calculated over a ball of finite radius $R_{\text{max}}$;
\item $f$ is sampled over a discrete domain and, thus, introducing related limitations, such as the risk of aliasing \citep{Pauley1990};
\item the distribution of the sampling points is usually irregular and non-uniform leading to larger errors where a lower sample density is present \citep{Smith1986, Smith1984, Buzzi1991, Barnes1994a} or in proximity to the domain boundaries \citep{Achtemeier1986};
\item an error is introduced by the back-interpolation function, $\phi$, from the Cartesian grid, $\mathbf{r_i}$, to the location of the samples, $\mathbf{r_j}$ (Eq. (\ref{eqn:Barne_scheme})) \citep{Pauley1990}. 
\end{itemize}

Before proceeding with further analysis, it is necessary to address the applicability of the LiSBOA to anisotropic and multi-chromatic scalar fields. Generally, the application of the LiSBOA with an isotropic weighting function is not recommended in case of severe anisotropy of the field and/or the data distribution. At the early stages of objective analysis techniques, the use of an anisotropic weighting function was proved to be beneficial to increase accuracy while highlighting patterns elongated along a specific direction, based on empirical \citep{Endlich1968} and theoretical arguments \citep{Sasaki1971}. Furthermore, the adoption of a directional smoothing parameter, $\sigma_p$, where $p$ is a generic direction, allows maximizing the utilization of the data retrieved through inherently anisotropic measurements, such as the line-of-sight fields detected by remote sensing instruments \citep{Askelson2000,Trapp2000}. With this in mind, we propose a linear scaling of the physical coordinates before the application of the LiSBOA to recover a pseudo-isotropic velocity field. The scaling reads as:
\begin{equation}\label{eqn:Coord_scaling}
	\tilde{x}_p=\frac{x_p-x_p^*}{\Delta n_{0,p}},
\end{equation}
where $\mathbf{x^*}$ is the origin of the scaled reference frame and $\Delta n_{0,p}$ is the scaling factor for the $p$-th direction.  Hereinafter, $\tilde{\cdot}$ refers to the scaled frame of reference. From a physical standpoint, the scaling is equivalent to the adoption of an anisotropic weighting function, while the re-scaling approach is preferred to ensure generality to the mathematical formulation outlined in this section. 

The scaling factor $\mathbf{\upDelta n_{0}}$ is an important parameter in the present framework and is referred to as fundamental half-wavelength, while the associated Fourier mode is denoted as the fundamental mode. The selection of the fundamental half-wavelength should be guided by a priori knowledge of the dominant length-scales of the flow in various directions. Modes exhibiting degrees of anisotropy different than that of the selected fundamental mode, will not be isotropic in the scaled mapping, which leads two consequences: first, their response will not be optimal, in the sense that the shortest directional wavelength can produce excessive damping of the specific mode \citep{Askelson2000}; second, the shape preservation of such non-spherical features in the field reconstructed through the LiSBOA is not ensured \citep{Trapp2000}.

Regarding the reconstruction of the flow statistics through the LiSBOA, two categories of error can be identified. The first is the statistical error due to the finite number of samples of the scalar field, $f$, available in time. This error is strictly connected with the local turbulence statistics, the sampling rate, and the duration of the experiment. The second error category is the spatial sampling error, which is due to the discrete sampling of $f$ in the spatial domain $\mathbf{x}$.
The Petersen-Middleton theorem \citep{Petersen1962} states that the reconstruction of a continuous and band-limited signal from its samples is possible if and only if the spacing of the sampling points is small enough to ensure non-overlapping of the spectrum of the signal with his replicas distributed over the so-called reciprocal lattice (or grid). The latter is defined as the Fourier transform of the specific sampling lattice. The 1D version of this theorem is the well-known Shannon-Nyquist theorem \citep{Shannon1984}. An application of this theorem to non-uniformly distributed samples, like those measured by remote sensing instruments, is unfeasible due to the lack of periodicity of the sampling points. To circumvent this issue, we adopted the approach suggested by \cite{Koch1983}, who defined the random data spacing, $\Delta d$, as the equivalent distance that a certain number of samples enclosed in a certain region, $N_{\text{exp}}$,  would have if they were uniformly distributed over a structured Cartesian grid. The generalized form of the random data spacing reads:
\begin{equation}\label{eq:random_data_spacing}
\Delta d(\mathbf{r_i})= \frac{V^\frac{1}{N}}{N_{\text{exp}}(\mathbf{r_i})^\frac{1}{N}-1},
\end{equation}
where $V$ is the volume of the hyper-sphere with radius $R_{\text{max}}=3 \sigma$ centered at the specific grid point and $N_{\text{exp}}$ represents the number of not co-located sample locations included within the hyper-sphere. Then, the Petersen-Middleton theorem for the reconstruction of the generic Fourier mode of half-wavelength $\mathbf{\upDelta n}$ can be translated as the following constraint:
\begin{equation}\label{eq:PM_thereom}
    \Delta d (\mathbf{r_i})< \Delta n_p,  ~ p =1,2, ..., N. 
\end{equation}
Violation of the inequality (\ref{eq:PM_thereom}) will lead to local aliasing, with the energy content of the under-sampled wavelengths being added to the low-frequency part of the spectrum. 

\begin{figure}[h]
\centerline{\includegraphics[width=\textwidth]{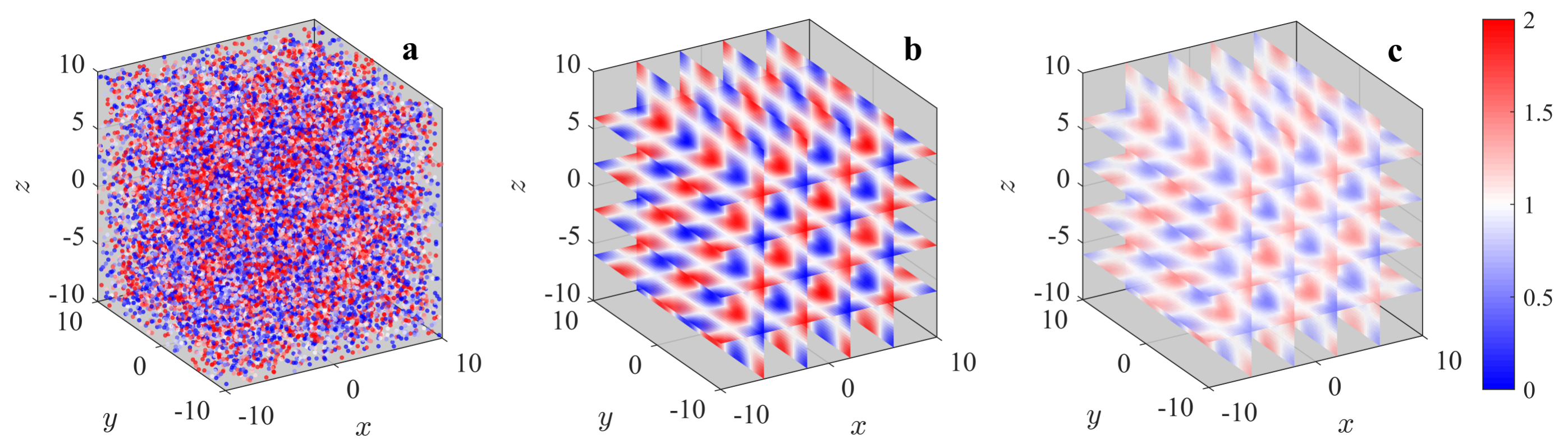}}
 \caption{Visualization of the LiSBOA applied to a Monte Carlo simulation of the synthetic field in Eq. (\ref{eqn:MC_field}) for the case with $N_s=20,000$, $L=200$, $\Delta n /\sigma =4$ and $m = 5$: a) samples; b) 3D reconstructed mean field, $g^m$; c) 3D reconstructed variance, $v^m$.}\label{fig:MC_200L_example}
\end{figure}

\begin{figure}[h]
\centerline{\includegraphics[width=\textwidth]{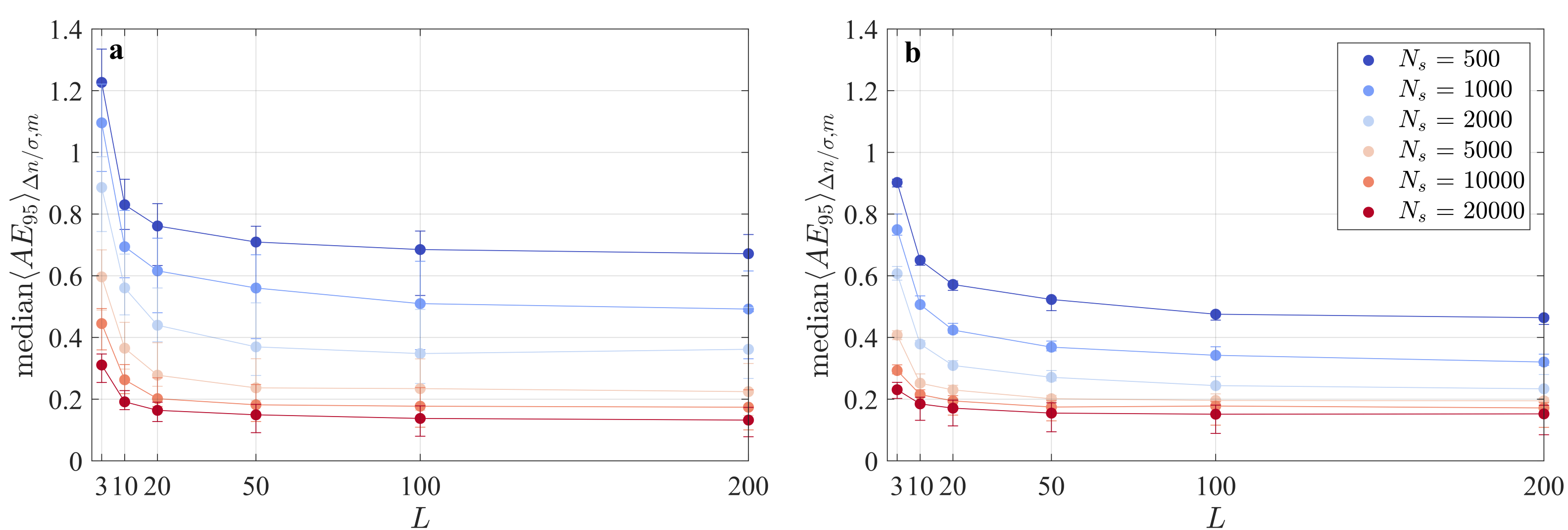}}
 \caption{Median of the $AE_{95}$ for all the tested half-wavelengths, $\Delta n/\sigma$, and the number of iterations, $m$: a) $AE_{95}$ of the mean field, $g^m$; b) $AE_{95}$ of the variance field, $v^m$. The error bars span the interquartile range.}\label{fig:MC_error_all}
\end{figure}

\begin{section}{LiSBOA assessment through Monte Carlo simulations}\label{sec:MonteCarlo}
The spectral response of the LiSBOA is studied through the Monte Carlo method. The goal of the present section is twofold: validating the analytical response of mean and variance (Eq. (\ref{eqn:Barnes_response})) and characterizing the sampling error of the LiSBOA as a function of the random data spacing. For these aims, a synthetic 3D scalar field is generated, while its temporal variability is reproduced locally by randomly sampling a normal probability density function. Specifically, the synthetic scalar field is: 
\begin{equation} \label{eqn:MC_field}
    f=\left[1 + \text{sin}\left(\frac{\pi}{\Delta n}x\right)\text{sin}\left(\frac{\pi}{\Delta n}y\right) \text{sin}\left(\frac{\pi}{\Delta n}z\right)  \right] +\left[1 + \text{sin}\left(\frac{\pi}{\Delta n}x\right) \text{sin}\left(\frac{\pi}{\Delta n}y\right) \text{sin}\left(\frac{\pi}{\Delta n}z\right)\right]^{0.5} \aleph(0,1),
\end{equation}
where $\aleph$ is a generator of random numbers with normal probability density function with mean value $0$ and standard deviation equal to $1$. The constant $1$ in the two terms on the RHS of Eq. (\ref{eqn:MC_field}) does not affect the LiSBOA response and is introduced to obtain both mean and variance of $f$ equal to the following function:
\begin{equation} \label{eq:fmean}
    \overline{f}=1 + \text{sin}\left(\frac{\pi}{\Delta n}x\right)\text{sin}\left(\frac{\pi}{\Delta n}y\right) \text{sin}\left(\frac{\pi}{\Delta n}z\right).
\end{equation}
It is noteworthy that $\overline{f}-1$ is a monochromatic isotropic function. 

An experimental sampling process is mimicked by evaluating the scalar field $f$ through randomly and uniformly distributed samples collected at the locations $\mathbf{r_j}$. The latter are distributed within a cube spanning the range $\pm 10\sigma$ in the three Cartesian directions. The total number of sampling points considered for each realization, $N_s$, is varied from 500 up to 20,000 to explore the effects of the sample density on the error. The sampling process is repeated $L$ times for each given distribution of $N_s$ points to capture the variability in the field introduced by the operator $\aleph$. The whole procedure can be considered as an idealized LiDAR experiment where a scan including $N_s$ sampling points is performed $L$ times to probe an ergodic turbulent velocity field.

Since the response is only a function of $\Delta n/\sigma$ and $m$ (Eq. (\ref{eqn:Barnes_response})), for the spectral characterization of the LiSBOA, the parameter $\Delta n/\sigma$ is varied among the following values: [1, 2, 3, 4, 5].  
An implementation of the LiSBOA algorithm for discrete samples is then applied to reconstruct the mean $g^m$ and variance $v^m$ of the scalar field $f$ over a Cartesian structured grid, $\mathbf{r_i}$, with a resolution of 0.25. Fig. \ref{fig:MC_200L_example} depicts an example of the reconstruction of the mean scalar field, $g^m$, and its variance, $v^m$, from the Monte Carlo synthetic dataset. 

For the error quantification, the 95-th percentile of the absolute error calculated at each grid point $\mathbf{r_i}$ ($AE_{95}$ hereinafter) is adopted:
\begin{equation}\label{eqn:AE95}
    AE_{95}(\Delta n/\sigma,m,N_s,L)=
    \begin{cases}
    \text{percentile}_{95}\langle|(g^{m}-1)-D^m(\cdot\overline{f}-1)| \rangle_{\mathbf{r_i}} & \text{for the mean} \\
    \text{percentile}_{95}\langle|(v^{m}-1)-D^0(\cdot\overline{f}-1)| \rangle_{\mathbf{r_i}} & \text{for the variance.} \\
    \end{cases}
\end{equation}
 The $AE_{95}$ quantifies the discrepancy between the outcome of the LiSBOA and the analytical input damped by the theoretical response evaluated over the Cartesian grid. As highlighted in Eq. (\ref{eqn:AE95}), the expected value of $AE_{95}$ is a function of the half-wavelength over the smoothing parameter, $\Delta n/\sigma$, the number of iterations, $m$, the number of samples, $N_s$, and the number of realizations, $L$. To investigate the link between $AE_{95}$ and the above-mentioned parameters, the Pearson correlation coefficients are analyzed (Table \ref{tab:AE_correlation}). The number of samples $N_s$, which is inversely proportional to the data spacing $\Delta d$ (Eq. \ref{eq:random_data_spacing}), is the variable exhibiting the strongest correlation with the error for both mean and variance. This indicates, as expected, that a larger number of samples for each measurement realization is always beneficial for the estimates of the statistics of the scalar field, $f$. Furthermore, the negative sign of correlations $\rho(AE_{95},N_s)$ and $\rho(AE_{95},\Delta n/\sigma)$, corroborate the hypothesis that the ratio $\Delta d/\Delta n$, i.e. the number of samples per half-wavelength, is the main driving factor for the sampling error \citep{Koch1983,Barnes1994a,Caracena1984}. 

The small positive correlation $\rho(AE_{95},m)$ detected for the mean is due to an amplification of the error occurring during the iterative process \citep{Barnes1964}. The issue will be discussed more in detail in section \ref{sec:Guidelines}. For the variance, $\rho(AE_{95},m)$ is practically negligible, confirming that the response of the higher-order statistics is insensitive to the number of iterations, $m$. Finally, the negative correlations with $L$ show that the statistical error is inversely proportional to the number of realizations collected. The dependence $\rho(AE_{95},L)$ is mainly due to the statistical error connected with the temporal sampling and, thus, the number of realizations, $L$, is progressively increased until convergence of the $AE_{95}$ is achieved. Fig. \ref{fig:MC_error_all} displays the behavior of the error as a function of $N_s$ and $L$. The values displayed represent the median for all the wavelength and iterations, being the $AE_{95}$ just mildly dependent on these parameters. As Fig. \ref{fig:MC_error_all} shows, increasing the number of realizations, $L$, beyond 100 has a negligible effect on the error.

To verify the analytical response of mean and variance of the scalar field, $f$, a numerical estimator of the response is defined as the median in space of the ratio between the field reconstructed via LiSBOA and the expected value of the synthetic input, as:
\begin{equation}\label{eq:MC_response}
    \text{}
    \begin{cases}
    D^m =  \text{median}\langle \frac{g^m -1}{\overline{f} -1} \rangle_{\mathbf{r_i}} & \text{for the mean} \\
    D^0 =\text{median}\langle \frac{v^m -1}{\overline{f}-1} \rangle_{\mathbf{r_i}} & \text{for the variance.} \\
    \end{cases}
\end{equation}
In the calculation of the numerical response through Eq. (\ref{eq:MC_response}), the influence of the edges is removed by rejecting points closer than $R_{\text{max}}$ to the boundaries of the numerical domain. Furthermore, the zero-crossings of the synthetic sine function ($|\overline{f}-1| < 0.1$) are excluded to avoid singularities. A comparison between the actual and the theoretical response (Eq. (\ref{eqn:Barnes_response})) for several wavelengths of the input function is reported in Fig. \ref{fig:MC_200L_resp} for the case with the highest number of samples $N_s=20,000$. An excellent agreement is observed between the theoretical prediction and the Monte Carlo outcome, which indicates that in the limit of negligible statistical error (large $L$) and adequate sampling (large $N_s$ and near-uniformly distributed samples) the response approaches the predictions obtained from the developed theoretical framework. 

The trend of the response of the mean (Fig. \ref{fig:MC_200L_resp}a) suggests that, for a given wavelength, the same response can be achieved for an infinite number of combinations $\sigma - m$ and, specifically, a larger $\sigma$ requires a larger number of iterations, $m$, to achieve a certain response $D^m$. It is noteworthy that for a smaller number of iterations, $m$, the slope of the response function is lower. This feature can be beneficial for practical applications for which the LiSBOA response will have small changes for small variations of $\Delta n$. However, a lower slope of the response function can be disadvantageous for short-wavelength noise suppression. Fig. \ref{fig:MC_200L_resp}b confirms that the response of the variance, and similarly for higher-order statistics, is not a function of the total number of iterations, $m$, and is equal to the response of the mean for the 0-th iteration, $D^0$.
 
Finally, the link between error and the random data spacing, $\Delta d$, is investigated. In Fig. \ref{fig:MC_200L_err}, the discrepancy with respect to theory quantified by the $AE_{95}$ is plotted versus the random data spacing normalized by the half-wavelength for a fixed total number of iterations $m=5$. The values displayed on the x-axis represent the median over all grid points, $\mathbf{r_i}$. This analysis reveals a strong correlation between the normalized random data spacing and the error. 
This analysis corroborates that, in the limit of negligible statistical error (viz. a high number of realizations, $L$), uncertainty is mainly driven by the local data density normalized by the wavelength, which is related to the Petersen-Middleton criterion. Indeed, the cases satisfying the Petersen-Middleton constraint (Eq. (\ref{eq:PM_thereom}))
are those exhibiting an $AE_{95}$ smaller than $\sim40\%$ of the amplitude of the harmonic function $\overline{f}$ for both mean and variance.  However, if a smaller error is needed, it will be necessary to reduce the maximum threshold value for  $\Delta d/\Delta n$.
\end{section}

\section{Guidelines for an efficient application of the LiSBOA to wind LiDAR data} \label{sec:Guidelines}
An efficient application of the LiSBOA to LiDAR data relies on the appropriate selection of the parameters of the algorithm, namely the fundamental half-wavelengths, $\mathbf{\upDelta n_0}$, the smoothing parameter, $\sigma$, the number of iterations, $m$, and the spatial discretization of the Cartesian grid, $\mathbf{dx}$. Furthermore, the data collection strategy must be designed to ensure adequate sampling of the spatial wavelengths of interest, so that the Petersen-Middleton constraint (Eq. (\ref{eq:PM_thereom})) is satisfied. In this section, we show that the underpinning theory of the LiSBOA, along with an estimate of the fundamental properties of the flow under investigation, can guide the  optimal design of a LiDAR experiment and evaluation of the statistics for a turbulent ergodic flow. The whole procedure can be divided into three phases: characterization of the flow, design of the experiment, and reconstruction of the statistics from the collected dataset. 

Firstly, it is crucial to estimate the integral quantities of the flow under investigation required for the application of the LiSBOA, such as extension of the spatial domain of interest, characteristic length-scales, integral time-scale, $\tau$, characteristic temporal standard deviation of the velocity, $\sqrt{\overline{u'^2}}$, and expected total sampling time, $T$, which depends on the typical duration of stationary boundary conditions over the domain. These estimates can be based on previous studies available in literature, numerical simulations, or preliminary measurements.

\begin{figure}[h]
\centerline{\includegraphics[width=\textwidth]{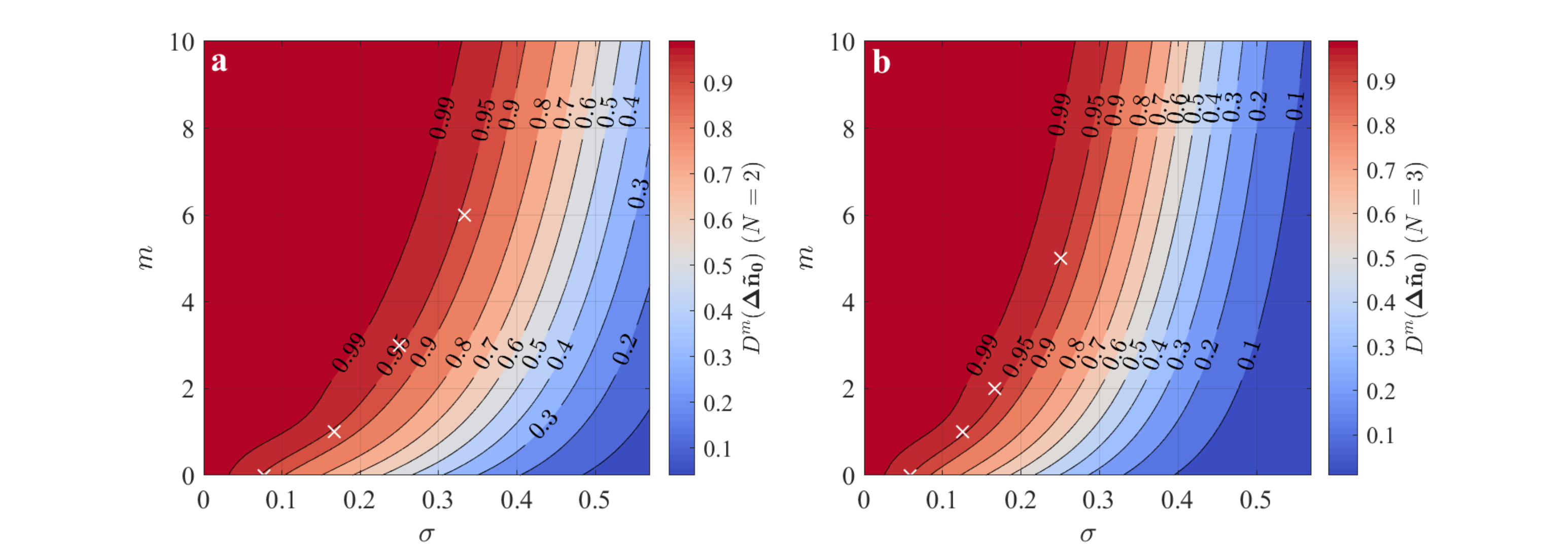}}
 \caption{Response of the fundamental mode in the scaled coordinates as a function of the number of iterations and the smoothing parameter: a) two-dimensional LiSBOA; b) three-dimensional LiSBOA. The white crosses indicate the pairs $\sigma - m$ provided in Table \ref{tab:Response_combinations}. }\label{fig:Response}
\end{figure}

Then, it is necessary to define the fundamental half-wavelengths, $\mathbf{\upDelta n_{0}}$, which are required for the coordinate scaling (Eq. (\ref{eqn:Coord_scaling})). It is advisable to impose the fundamental half-wavelengths equal to (or even smaller than) the estimated characteristic length-scales of the smallest spatial features of interest in the flow. This ensures isotropy of the mode associated with the fundamental half-wavelength (and all the modes characterized by the same degree of anisotropy) and guides the selection of the main input parameters of the LiSBOA algorithm, i.e. smoothing parameter, $\sigma$, and number of iterations, $m$.
Indeed, $\mathbf{\upDelta n_{0}}$ can be considered as the cut-off half-wavelength of the spatial low-pass filter represented by the LiSBOA operator. To this aim, it is necessary to select $\sigma$ and $m$ to obtain a response of the mean associated with the fundamental mode, $D^m(\mathbf{\upDelta \tilde{n}_0})$, as close as possible to 1. After the coordinate scaling (Eq. (\ref{eqn:Coord_scaling})), the response of the fundamental mode is universal and it is reported in Fig. \ref{fig:Response}. For instance, if we select a response equal to 0.95, then all the points lying on the iso-contour defined by the equality  $D^m(\mathbf{\upDelta \tilde{n}_0})=0.95$ give, in theory, the same response for the mean of the scalar field $f$. This implies that an infinite number of combinations $\sigma - m$ allow obtaining a response of the mean equal to the selected value.  However, with increasing $\sigma$ the response at the 0-th iteration, $D^0(\mathbf{\upDelta \tilde{n}_0})$, reduces, which indicates a lower response for higher-order statistics. For the LiSBOA application, the following aspects should be also considered:
\begin{itemize}
    \item the smaller $\sigma$, the smaller the radius of influence of the LiSBOA, $R_{\text{max}}$, and, thus, the lower the number of samples averaged per grid node, $N_{\text{exp}}$, and the greater the statistical uncertainty;
    \item an excessively large $m$ can lead to overfitting of the experimental data and noise amplification \citep{Barnes1964};
    \item the higher $m$, the higher the slope of the response function (see Fig. \ref{fig:MC_200L_resp}), which improves the damping of high-frequency noise, but it produces a larger variation of the response of the mean with different spatial wavelengths;
    \item the radius of influence $R_{\text{max}}$ (and therefore $\sigma$) can affect the data spacing $\Delta d$ in case of non-uniform data distribution.
\end{itemize}
Few handy combinations of smoothing parameter and total iterations for $D^m(\mathbf{\upDelta \tilde{n}_0})=0.95$ are provided in Table \ref{tab:Response_combinations}. As mentioned above, all these $\sigma-m$ pairs allow achieving roughly the same response for the mean, while the response for the higher-order statistics reduces with an increasing number of iterations, $m$.

As a final remark about the selection of $\mathbf{\upDelta n_0}$, we should consider that, if the fundamental half-wavelength is too large compared to the dominant modes in the flow, small-scale spatial oscillations of $f$ will be smoothed out during the calculation of the mean, with consequent underestimated gradients and incorrect estimates of the high-order statistics due to the dispersive stresses \citep{Arenas2019}. On the other hand, the selection of an overly small $\mathbf{\upDelta n_0}$, would require an excessively fine data spacing to satisfy the Petersen-Middleton constraint (Eq. (\ref{eq:PM_thereom})), which may lead to an overly long sampling time or even exceed the sampling capabilities of the LiDAR.

The optimal LiDAR scanning strategy aimed to characterize atmospheric turbulent flows implies finding a trade-off between a sufficiently fine data spacing, which is quantified through $\Delta d$ in the present work (Eq.(\ref{eq:random_data_spacing})), and an adequate number of time realizations, $L$, to reduce the temporal statistical uncertainty. Considering a total sampling period $T$, for which statistical stationarity can be assumed, and a pulsed LiDAR that scans $N_r$ points evenly spaced along the LiDAR laser beam, with a range gate $\Delta r$ and accumulation time $\tau_a$, the total number of collected velocity samples is then equal to $N_s= N_r\cdot T/\tau_a$. The angular resolution of the LiDAR scanning head, $\Delta \theta$, can be selected to modify the angular spacing between consecutive line-of-sights (i.e. the data spacing) and the total sampling period for a single scan, $\tau_s$ (i.e. the number of realizations, $L$). 

The design of a LiDAR scan aiming to reconstruct turbulent statistics of an ergodic flow thorough LiSBOA can be formalized as a two-objective (or Pareto front) optimization problem. The first cost function of the Pareto front, which is referred to as $\epsilon^I$, is the percentage of grid nodes for which the Petersen-Middleton constraint applied to the smallest half-wavelength of interest (i.e. $\mathbf{\upDelta n_0}$), is not satisfied. Regarding the scaled reference frame, this can be formalized as:
\begin{equation}\label{eq:epsilon1}
    \epsilon^I(\Delta \theta, \sigma)=\frac{\sum_{i=1}^{N_i}[\Delta \tilde{d}>1]}{N_i},
\end{equation}
where the square brackets are Iverson brackets and $N_i$ is the total number of nodes in the Cartesian grid, $\mathbf{r_i}$. For a more conservative formulation, it is recommended to reject all the points with a distance smaller than $R_{\text{max}}$ from an under-sampled grid node, i.e. with $\Delta \tilde{d}>1$. This condition will ensure that the statistics are based solely on regions that are adequately sampled. The cost function $\epsilon^{I}$ depends not only on the angular resolution but also on $R_{\text{max}}$, which is equal to $3\sigma$ in this work. In general, increasing $\sigma$ results in a larger number of samples considered for the calculation of the statistics at each grid point $\mathbf{r_i}$ and, thus, in a reduction of $\epsilon^I$. Therefore, a larger $\sigma$ entails a larger percentage of the spatial domain fulfilling the Petersen-Middleton constraint. The smoothing parameter, $\sigma$, also plays a fundamental role in the response of higher-order statistical moments. Specifically, if the reconstruction of the variance or higher-order statistics is important, the response $D^0(\mathbf{\upDelta \tilde{n}_0})$ should be included in the Pareto front analysis as an additional constraint.

\begin{figure}[h]
\centerline{\includegraphics[width=0.5\textwidth]{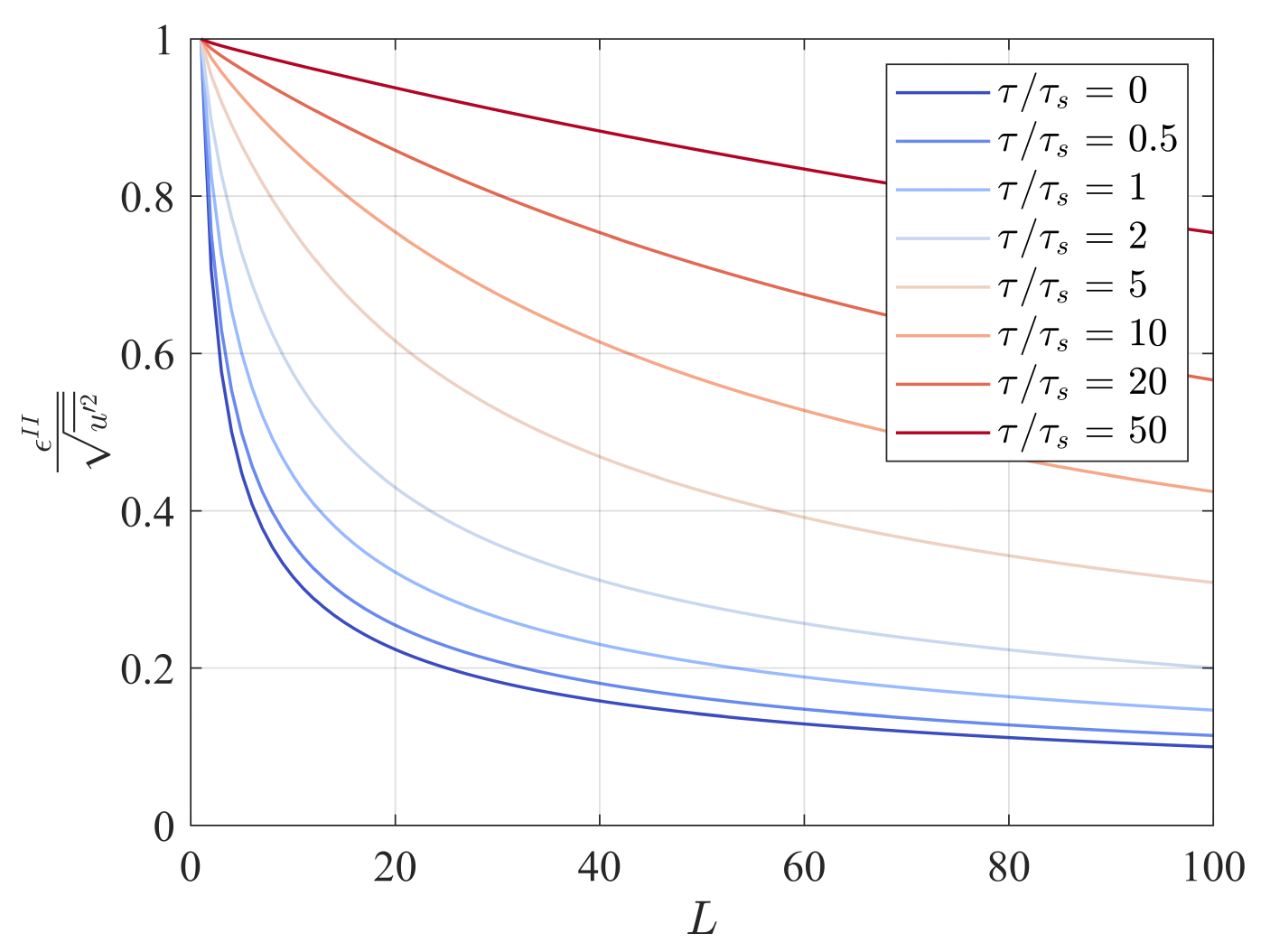}}
 \caption{Standard deviation of the sample mean normalized by the standard deviation of velocity as a function of the number of realizations, $L$, and for different values of the ratio between the integral time-scale and the sampling time, $\tau/\tau_s$.}\label{fig:Error_mean}
\end{figure}

The second cost function for the optimal design of LiDAR scans, $\epsilon^{II}$, is equal to the standard deviation of the sample mean, which, for an autocorrelated signal, is \citep{Bayley1946}:
\begin{equation}\label{eq:Error_mean}
    \epsilon^{II}(\Delta \theta) =  \sqrt{\overline{u'^2}}\sqrt{ \frac{1}{L} + \frac{2}{L^2} \sum_{p=1}^{L-1} (L-p)~ \rho_{p}} \sim \sqrt{\overline{u'^2}}\sqrt{\frac{1}{L} + \frac{2}{L^2} \sum_{p=1}^{L-1} (L-p)~ e^{-\frac{ \tau_s}{\tau}p}},
\end{equation}
where $\rho_{p}$ is the autocorrelation function at lag $p$, $\tau$ is the integral time-scale and the approximation is based on \cite{George1978}. The velocity variance, $\sqrt{\overline{u'^2}}$, and the autocorrelation, $\rho_{p}$, are functions of space; however, to a good degree of approximation, they can be replaced by a representative value and considered as uniform in space. Fig. \ref{fig:Error_mean} shows the standard deviation of the sample mean normalized by the standard deviation of the velocity as a function of the number of realizations, $L$, and for different integral time-scales, $\tau$. It is noteworthy that the standard deviation of the sample mean represents the uncertainty of the time-average of each measurement point, $\mathbf{r_j}$, while the final uncertainty of the mean field at the grid nodes $\mathbf{r_i}$ is generally reduced due to the spatial averaging process intrinsic to the LiSBOA. 

\begin{figure}[h]
\centerline{\includegraphics[width=\textwidth]{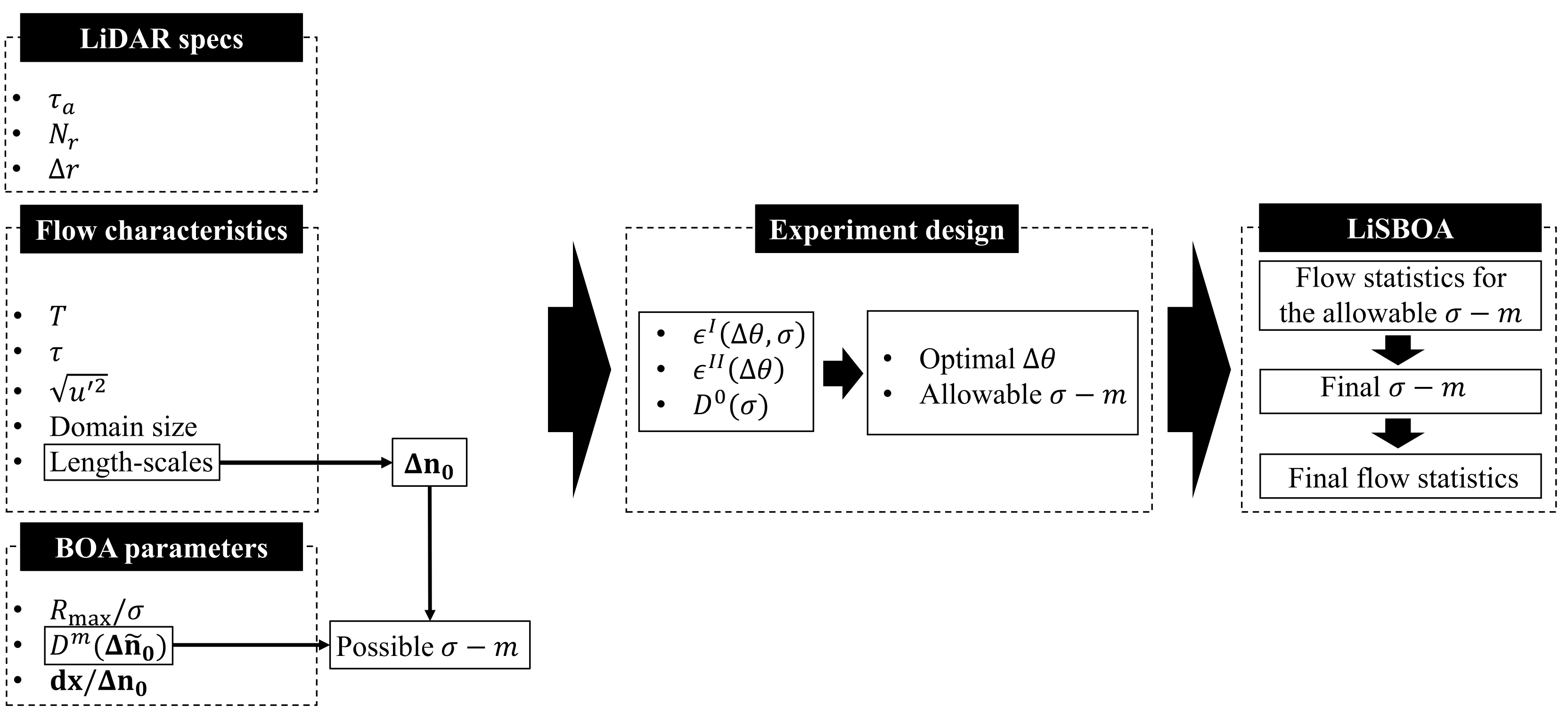}}
 \caption{Schematic of the LiSBOA procedure for the optimal design of LiDAR scans and reconstruction of the statistics for a turbulent ergodic flow. }\label{fig:FlowChart}
\end{figure}

The whole procedure for the design of a LiDAR scan and retrieval of the statistics is reported in the flow chart of Fig. \ref{fig:FlowChart}. Summarizing, from a preliminary analysis of the velocity field under investigation, we determine the maximum total sampling time, $T$, the characteristic integral time-scale, $\tau$, the characteristic velocity variance, $\overline{u'^2}$, the fundamental half-wavelengths $\mathbf{\upDelta n_0}$. This information, together with the settings of the LiDAR (namely, the accumulation time, $\tau_a$, the number of points per beam, $N_r$, and the gate length, $\Delta r$), allow for generating the Pareto front as a function of $\Delta \theta$ and for different values of $\sigma$. Based on the specific goals of the LiDAR campaign in terms of coverage of the selected domain (i.e. $\epsilon^{I}$), the statistical significance of the data (i.e. $\epsilon^{II}$) and, eventually, the response of the higher-order statistical moments (i.e. $D^0(\mathbf{\upDelta \tilde{n}_0})$), the LiSBOA user should select the optimal angular resolution, $\Delta \theta$, and the set of allowable $\sigma$ values. Due to the above-mentioned non-ideal effects on the LiSBOA, the selection of $\sigma$ is finalized during the post-processing phase when the LiDAR dataset is available and the statistics can be calculated for different pairs of $\sigma-m$ values.
For the resolution of the Cartesian grid, \cite{Koch1983} suggested that it should be chosen as a fraction of the data spacing, which, in turn, is linked to the fundamental half-wavelength. The same author suggested a grid spacing included in the range $\mathbf{d x}\in [\mathbf{\upDelta n_0}/3,\mathbf{\upDelta n_0}/2]$. In this work, we have used $\mathbf{d x} = \mathbf{\upDelta n_0}/4$, which ensures a good grid resolution with acceptable computational costs.

By following the steps outlined in the present section, the mean, variance, or even higher-order statistical moments of the velocity field can be accurately reconstructed for the wavelengths of interest. It is worth mentioning that the LiSBOA of wind LiDAR data should always be combined with a robust quality-control process of the raw measurements. Indeed, the space-time averaging operated by the LiSBOA makes the data analysis sensitive to the presence of data outliers, which need to be identified and rejected beforehand to prevent contamination of the final statistics.

\section{Conclusions}\label{sec:Conclusions}
A revisited Barnes objective analysis for sparse and non-uniformly distributed LiDAR data has been formulated to calculate mean, variance, and higher-order statistics of the wind velocity field over a structured N-dimensional Cartesian grid. This LiDAR Statistical Barnes Objective Analysis (LiSBOA) provides a theoretical framework to quantify the response in the reconstruction of the velocity statistics as a function of the spatial wavelengths of the velocity field under investigation and quantification of the sampling error.
The LiSBOA has been validated against volumetric synthetic 3D data generated through Monte-Carlo simulations. The results of this test have shown that the  sampling error for a monochromatic scalar field is mainly driven by the data spacing normalized by the half-wavelength. 

Guidelines for the optimal design of scans performed with a scanning Doppler pulsed wind LiDAR and calculation of wind velocity statistics have been provided by leveraging the LiSBOA. The optimization problem consists in providing background information about the turbulent flow under investigation, such as expected velocity variance and integral length scales, technical specifications of the LiDAR, such as range gate and accumulation time, and spatial wavelengths of interest for the velocity field. The formulated optimization problem has two cost functions, namely the percentage of grid nodes not satisfying the Petersen-Middleton constraint for the smallest half-wavelength of interest (i.e. lacking adequate spatial resolution to avoid aliasing in the statistics), and the standard deviation of the sample mean. The output of the optimization problem are the LiDAR angular resolution and, for a given response of the mean field, the allowable smoothing parameters and number of iterations to use for the LiSBOA. 

In the companion paper \citep{PartII}, the LiSBOA is applied to velocity fields associated with wind turbine wakes obtained through large-eddy simulations and LiDAR measurements. As a final note, this work is intended to contribute to the improvement and standardization of the LiDAR data collection and analysis methodology. We believe that the formulation of validated tools for quantitative analysis of LiDAR data represents an important process to unleash the full potential of the scanning LiDAR technology for investigations of atmospheric turbulent flows.

%
%
\datastatement
The LiSBOA algorithm is implemented in a publicly available code available at https://www.utdallas.edu/windflux/.

\acknowledgments
This research has been funded by a grant from the National Science Foundation CBET Fluid Dynamics, award number 1705837. This material is based upon work supported by the National Science Foundation under grant IIP-1362022 (Collaborative Research: I/UCRC for Wind Energy, Science, Technology, and Research) and from the WindSTAR I/UCRC Members: Aquanis, Inc., EDP Renewables, Bachmann Electronic Corp., GE Energy, Huntsman, Hexion, Leeward Asset Management, LLC, Pattern Energy, EPRI, LM Wind, Texas Wind Tower and TPI Composites. Any opinions, findings, and conclusions or recommendations expressed in this material are those of the authors and do not necessarily reflect the views of the sponsors. Texas Advanced Computing Center is acknowledged for providing computational resources.



 \appendix

\appendixtitle{Derivation of the analytical response function of the LiSBOA}

The first iteration of the LiSBOA produces a weighted average in space of the scalar field, $f$, with the weights being Gaussian functions centered at the specific grid nodes, $\mathbf{x}$. In the limit of a continuous function defined over an infinite domain, Eq. (\ref{eqn:Theory_Barnes}) represents the convolution between the scalar field, $f$, and the Gaussian weights, $w$. Therefore, the response function of the LiSBOA, can be expressed in the spectral domain as \citep{Pauley1990}:

\begin{equation}
    D^0=\frac{\mathfrak{F}[g^0]}{\mathfrak{F}[f]}=\mathfrak{F}[w],
    \label{eqn:Theory_Barnes_app}
\end{equation}
where the operator $\mathfrak{F}$ indicates the Fourier transform (FT). The FT of the weighting function in Eq. (\ref{eqn:Theory_Barnes_app}) can be conveniently recast as the product of N one-dimensional FT:
\begin{equation}\label{eq:FT_Barnes}
\mathfrak{F}[w]= \prod_{p=1}^N  \int_{-\infty}^\infty \frac{1}{\sqrt{2\pi}\sigma}e^ \frac{-x_p^2}{2 \sigma^2}\cdot e^{-\mathrm{i} k_p x_p}  d x_p,
\end{equation}
where $k_p$ is the directional wavenumber and $\mathrm{i}=\sqrt{-1}$. Hence, by leveraging the closed-form FT of the Gaussian function \citep{Greenberg1998}:
\begin{equation}\label{eqn:Gaussian_FT}
    \mathfrak{F}\left[\frac{1}{\sqrt{2\pi}\sigma} e^\frac{-x^2}{2\sigma^2}\right]=e^\frac{-k^2\sigma^2}{2}
\end{equation}
we get the desired results, i.e.:
\begin{equation}
 D^0(\textbf{k})=e^{-\frac{\sigma^2}{2} |\mathbf{k}|^2}.
\end{equation}

%
%
%

\bibliographystyle{ametsoc2014}
\bibliography{references2}

\begin{table}[h]
\caption{Pearson correlation coefficient between the $AE_{95}$ of mean and variance and the parameters $\Delta n/\sigma$, $m$, $N_s$, $L$. The values between parenthesis represent the 95\% confidence bounds.}\label{tab:AE_correlation}
\begin{center}
\begin{tabular}{l c c c c }
\topline   & $\mathbf{\upDelta n/\sigma}$   & $\mathbf{m}$ & $\mathbf{N_s}$& $\mathbf{L}$ \\
\hline
\textbf{$\mathbf{AE_{95}}$ of mean} & -0.259 (-0.303, -0.210) &   0.257 (0.211, 0.301)  & -0.709 (-0.732, -0.684) &  -0.171 (-0.217, -0.124) \\ 
\textbf{$\mathbf{AE_{95}}$ of variance} &   -0.069 (-0.117, -0.021) &  -0.03 (-0.078, 0.019)  & -0.694 (-0.718, -0.668) &  -0.206 (-0.251, -0.159) \\
\hline
\end{tabular}
\end{center}
\end{table}

\begin{table}[h]
\caption{Selected combinations of $\sigma$ and $m$ to achieve a $\sim$95\% recovery of the mean of the selected fundamental half-wavelength and associated response of the higher-order moments (HOM).}\label{tab:Response_combinations}
\begin{center}
\begin{tabular}{lccc|lccc}
\hline
\hline
\multicolumn{4}{c|}{$\mathbf{N=2}$} &  \multicolumn{4}{c}{$\mathbf{N=3}$}\\

\hline $\sigma$   & $\mathbf{m}$&$\mathbf{D^m}$\textbf{ (mean)} & $\mathbf{D^0}$ \textbf{(HOM)} & $\sigma$   & $\mathbf{m}$&$\mathbf{D^m}$ \textbf{(mean) }& $\mathbf{D^0}$ \textbf{(HOM)}\\
 1/3    & 6   & 0.942 & 0.334 &  1/4                             & 5                 & 0.952  & 0.397 \\
  1/4     & 3   & 0.955  & 0.540 & 1/6                             & 2                 & 0.961  & 0.663 \\
 1/6     & 1   & 0.942  & 0.76 &1/8                              & 1                 & 0.957  & 0.793 \\
 1/13     & 0   & 0.943 & 0.943 & 1/17                            & 0                 & 0.950  & 0.950 \\
\hline
\end{tabular}
\end{center}
\end{table}


\begin{figure}[h]
\centerline{\includegraphics[width=\textwidth]{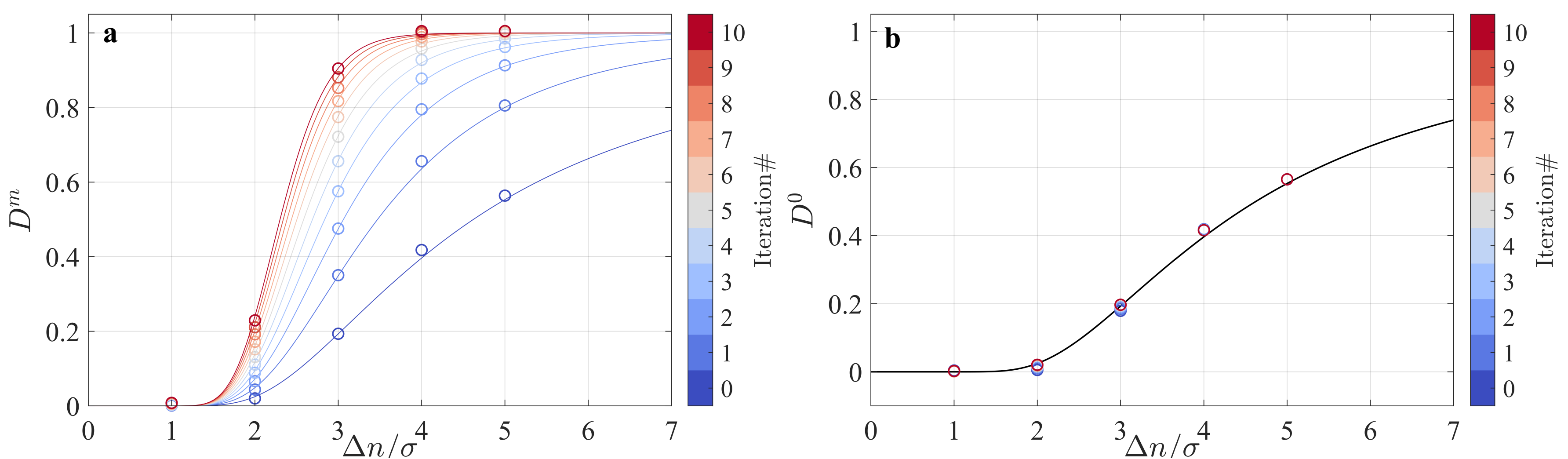}}
 \caption{Validation of the 3D theoretical response of the LiSBOA for the case $N_s=20000$ - $L=200$: a) mean; b) variance. The circles are the numerical output of the Monte Carlo simulation (Eq. (\ref{eq:MC_response})), while the continuous lines represent Eq. (\ref{eqn:Barnes_response}). }\label{fig:MC_200L_resp}
\end{figure}

\begin{figure}[h]
\centerline{\includegraphics[width=\textwidth]{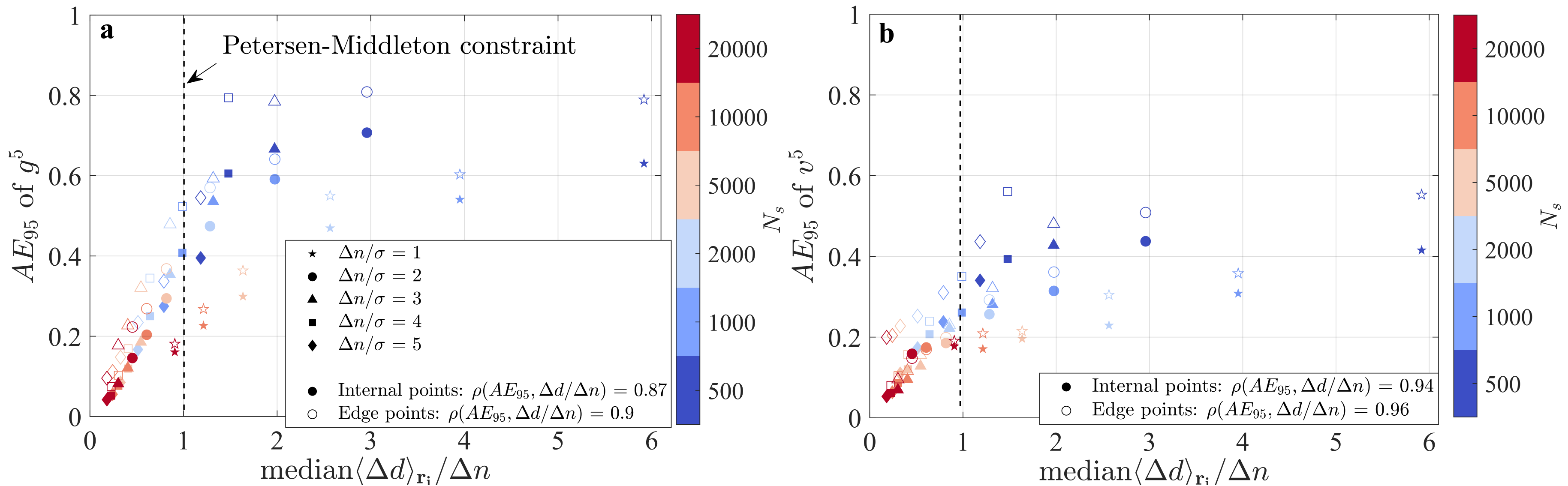}}
 \caption{$AE_{95}$ as a function of the random data spacing (Eq. (\ref{eq:random_data_spacing})) for the case with $m=5$, $N=20000$ and $L=200$: a) error on the mean; b) error on the variance. The full symbols refer to points not affected by the presence of the finite boundaries of the domain, while the empty symbols are taken within a distance of less than $R_{\text{max}}$ from the boundaries.}\label{fig:MC_200L_err}
\end{figure}

\end{document}